\documentclass[sigconf]{acmart} 

\usepackage{subcaption}
\usepackage{url}
\usepackage{enumitem}
\usepackage{multirow}
\usepackage{array}      
\usepackage{siunitx}
\usepackage{booktabs}   
\usepackage{siunitx}    
\usepackage{graphicx}
\usepackage{comment}
\usepackage[table]{xcolor}
\usepackage{amsmath}
\usepackage{pgfplots}
\pgfplotsset{compat=1.18}
\usepackage{xcolor}

\newcommand{\etal}{\emph{et al.}}
\newcommand{\eg}{\emph{e.g.}}
\newcommand{\ie}{\emph{i.e.}}

\newif\ifshownotes
\shownotesfalse   

\ifshownotes
  \newcommand{\zznote}[1]{\textcolor{magenta}{ZZ:#1}}
  \newcommand{\sfnote}[1]{\textcolor{orange}{SF:#1}}
\else
  \newcommand{\zznote}[1]{}
  \newcommand{\sfnote}[1]{}
\fi
\AtBeginDocument{%
  }

\setcopyright{acmlicensed}
\copyrightyear{2026}
\acmYear{2026}
\setcopyright{cc}
\setcctype{by}
\acmConference[SIGIR '26]{Proceedings of the 49th International ACM SIGIR Conference on Research and Development in Information Retrieval}{July 20--24, 2026}{Melbourne, VIC, Australia}
\acmBooktitle{Proceedings of the 49th International ACM SIGIR Conference on Research and Development in Information Retrieval (SIGIR '26), July 20--24, 2026, Melbourne, VIC, Australia}
\acmDOI{10.1145/3805712.3809598}
\acmISBN{979-8-4007-2599-9/2026/07}

\begin{document}

\title{Pretrain-then-Adapt: Uncertainty-Aware Test-Time Adaptation \\ for Text-based Person Search}


\author{Jiahao Zhang}
\affiliation{
  \institution{University of Macau}
  \city{Macau}
  \country{China}
}
\email{yc57963@um.edu.mo}

\author{Shaofei Huang}
\affiliation{
  \institution{University of Macau}
  \city{Macau}
  \country{China}
}
\email{nowherespyfly@gmail.com}

\author{Yaxiong Wang}
\affiliation{
  \institution{Hefei University of Technology}
  \city{Hefei}
  \country{China}
}
\email{wangyx@hfut.edu.cn}

\author{Zhedong Zheng}
\authornote{Corresponding author.}
\affiliation{
  \institution{University of Macau}
  \city{Macau}
  \country{China}
}
\email{zhedongzheng@um.edu.mo}

\renewcommand{\shortauthors}{Zhang et al.}

\begin{abstract}

Text-based person search faces inherent limitations due to data scarcity, driven by stringent privacy constraints and the high cost of manual annotation. To mitigate this, existing methods usually rely on a \textbf{Pretrain-then-Finetune} paradigm, where models are first pretrained on synthetic person-caption data to establish cross-modal alignment, followed by fine-tuning on labeled real-world datasets. However, this paradigm lacks practicality in real-world deployment scenarios, where large-scale annotated target-domain data is typically inaccessible. 
In this work, we propose a new \textbf{Pretrain-then-Adapt} paradigm that eliminates reliance on extensive target-domain supervision through an offline test-time adaptation manner, enabling dynamic model adaptation using only unlabeled test data with minimal post-train time cost.
To mitigate overconfidence with false positives of previous entropy-based test-time adaptation, we propose an Uncertainty-Aware Test-Time Adaptation (UATTA) framework, which introduces a bidirectional retrieval disagreement mechanism to estimate uncertainty, \ie, low uncertainty is assigned when an image-text pair ranks highly in both image-to-text and text-to-image retrieval, indicating high alignment; otherwise, high uncertainty is detected. 
This indicator drives offline test-time model recalibration without labels, effectively mitigating domain shift. 
We validate UATTA on four benchmarks, \ie,  CUHK-PEDES, ICFG-PEDES, RSTPReid, and PAB, showing consistent improvements across both CLIP-based (one-stage) and XVLM-based (two-stage) frameworks. 
Ablation studies confirm that UATTA outperforms existing offline test-time adaptation strategies, establishing a new benchmark for label-efficient, deployable person search systems. Our code is available at \url{https://github.com/nkuzjh/UATTA}.

\label{0_abstract}
\end{abstract}




\maketitle

\begin{figure}[!t]
    \centering
    \includegraphics[width=\columnwidth]{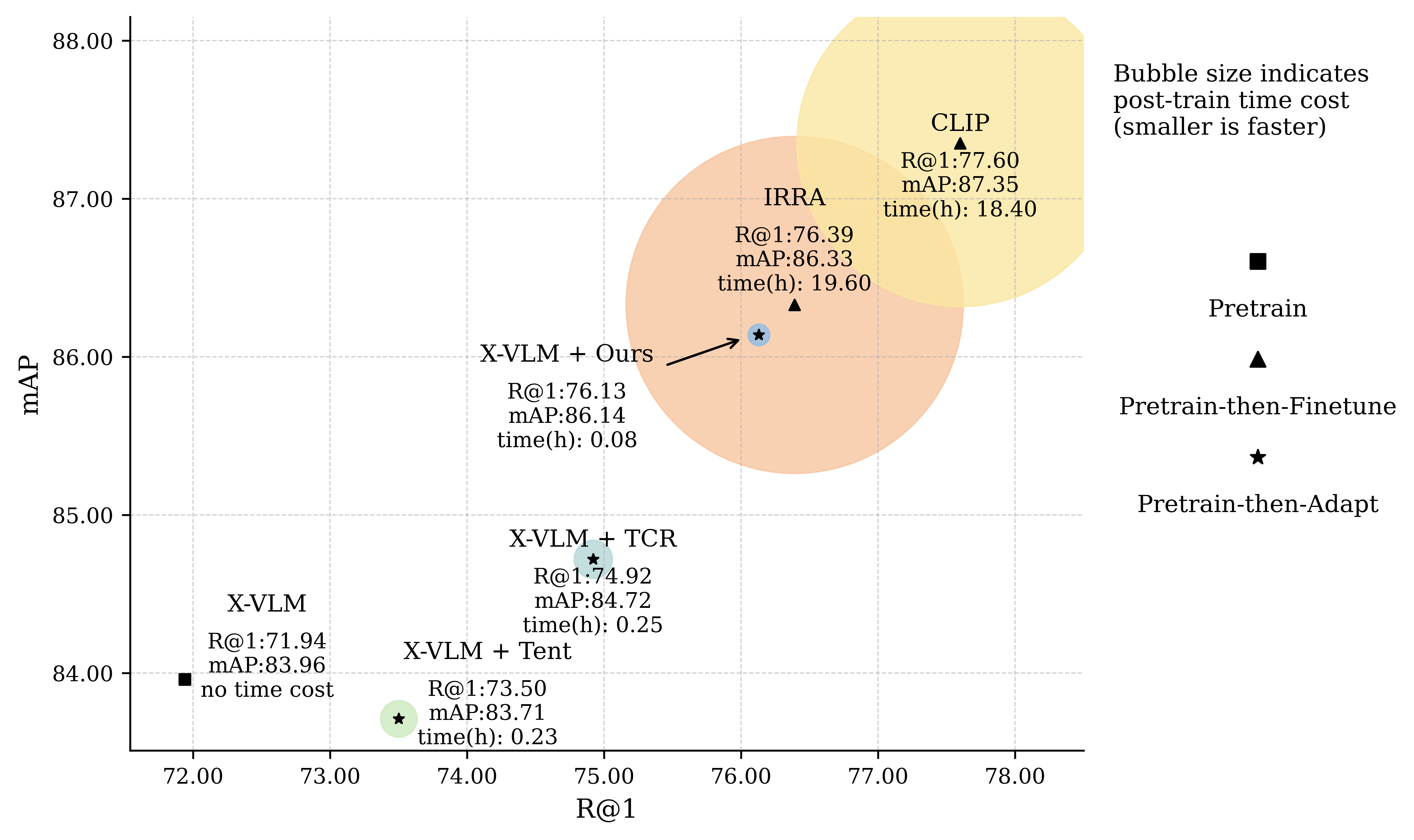}
    \vspace{-.3in}
    \caption{\textbf{Accuracy vs. efficiency trade-off on the PAB benchmark~\citep{yang2024beyond}.} Built upon a pretrained X-VLM~\citep{zeng2021multi} backbone, our method follows the \textbf{Pretrain-then-Adapt} paradigm and substantially improves retrieval performance with minimal adaptation cost (\textit{i.e.}, post-train GPU time). Compared with Pretrain-then-Finetune approaches, our method reduces post-train GPU time of adaptation by 99.6\% while achieving competitive performance, approaching or even matching strong Pretrain-then-Finetune methods such as IRRA~\citep{cvpr23crossmodal}.
    We present Pretrain-then-Finetune results (\ie, CLIP~\citep{radford2021learning} and IRRA~\citep{cvpr23crossmodal}) as an upper bound on achievable performance under extensive finetuning. In contrast, our method reaches a favorable operating point close to this upper bound with significantly lower post-train time cost, highlighting its improved accuracy vs. efficiency trade-off under the \textbf{Pretrain-then-Adapt} setting.}
    \label{fig0:acc_vs_spd}
    \vspace{-.2in}
\end{figure}

\section{Introduction}


Text-based person search~\citep{li2017person, zhu2021dssl, ding2021seman}, which involves matching natural language descriptions to specific individuals within large-scale image galleries, is a critical task with applications ranging from locating missing persons~\citep{bukhari2023language} to enhancing smart city management~\citep{khan2021deepreid, zheng20242}. Unlike conventional image-based person re-identification~\citep{zheng2015scalable, zheng2017unlabeled}, the incorporation of text modality offers a more intuitive and accessible query interface for system operators~\citep{zheng2020dual}.

Despite its practical advantages, the efficacy of current methods is severely hampered by the domain shift problem, where models trained in controlled settings exhibit significant performance degradation when deployed in unseen, real-world environments. 
State-of-the-art approaches typically attempt to mitigate this challenge through a \textbf{Pretrain-then-Finetune} paradigm~\citep{shao2023unified, cvpr23crossmodal, nguyen2024tackling, tan2024harnessing, jiang2025modeling}. This involves first pretraining on large-scale, often synthetic, person-caption datasets to establish preliminary cross-modal alignments, followed by fine-tuning on domain-specific annotated datasets such as CUHK-PEDES~\citep{li2017person}. However, the reliance on labeled data for fine-tuning renders this paradigm impractical for many real-world deployments. In practice, target domain labels are typically unavailable due to stringent privacy regulations~\citep{gaikwad2023real} and prohibitive annotation costs~\citep{shao2023unified}.

To address this limitation, we introduce the source-free offline test-time adaptation (TTA)~\citep{wang2021tent, dong2025domain} to the cross-modal retrieval task, formulating a \textbf{Pretrain-then-Adapt} paradigm, leveraging to adapt a pretrained model to a new target domain using only unlabeled test samples. Such strategy directly performs tailored adaptation to the specific data distribution of the current test set, therefore alleviating the reliance of labeled target domain data..
As depicted in Fig.~\ref{fig0:acc_vs_spd}, the Pretrain-then-Adapt paradigm shows superior efficiency and competitive performance compared to traditional Pretrain-then-Finetune paradigm, owing to its independence of fine-tuning on domain specific labeled data. Consequently, this paradigm demonstrates superior efficiency requiring orders-of-magnitude lower adaptation cost in contrast to traditional Pretrain-then-Finetune paradigm.
A prevailing practice within this paradigm involves adapting the model via entropy minimization~\citep{wang2021tent, yang2022test}, a TTA strategy widely adopted in image classification. By minimizing prediction entropy in an online or offline manner, the model is forced to sharpen its decision boundaries and increase its confidence in unlabeled target samples.
However, this approach presents a significant risk of error accumulation where the model can be overconfident in its own erroneous predictions, reinforcing them during adaptation and converging to a suboptimal state~\citep{zhao2024testtime}. In the context of retrieval, this implies that the model treats false positives as reliable as true positives, thereby amplifying the impact of wrong supervision signals.
This raises a crucial research question: \textit{How can we mitigate the risk of overconfident, erroneous adaptation in cross-modal person retrieval, a task that demands fine-grained matching?}

We argue that the key to mitigating the issue of overconfident false positives hinges on reliable uncertainty calibration. As illustrated in Fig.~\ref{fig2}, samples exhibiting high uncertainty are predominantly concentrated among false positives. This suggests that high uncertainty serves as an effective proxy for identifying false positives. Consequently, we introduce the \textbf{Uncertainty-Aware Test-Time Adaptation (UATTA)} framework, which leverages prediction uncertainty to re-calibrate the offline adaptation process on the entire test set. However, since legitimate uncertainty is intractable to estimate directly without ground truth, we propose \textit{bidirectional retrieval disagreement} as a tractable proxy. A high-uncertainty match will exhibit incongruity between the text-to-image and the corresponding image-to-text retrieval directions, whereas a confident, low-uncertainty match will show symmetric alignment. We provide a theoretical justification that this metric effectively gauges prediction uncertainty. This principle allows us to identify and down-weight potential false positives to avoid overconfidence during adaptation.

Specifically, for a one-stage retrieval model based on CLIP~\citep{radford2021learning}, we quantify bidirectional retrieval disagreement uncertainty using the relative disparity between mutual retrieval probabilities derived from the Image-Text Contrastive (ITC) loss. This bidirectional retrieval disagreement uncertainty measure is then used to rectify the entropy minimization objective by re-weighting it with the reciprocal of the uncertainty. 
For two-stage retrieval architectures like XVLM~\citep{zeng2021multi}, we apply the same principle to modulate the entropy of the fine-grained predictions from the Image-Text Matching (ITM) module. 
In both architectures, this uncertainty-aware rectification acts as a dynamic filter, effectively suppressing gradients from overconfident false positives to prevent error accumulation of vanilla entropy minimization. Consequently, by ensuring adaptation is solely based on reliable alignments, UATTA bridges the gap between unsupervised adaptation and supervised fine-tuning.
As shown in Fig.~\ref{fig0:acc_vs_spd}, our approach realizes a superior accuracy and efficiency trade-off, delivering performance competitive with expensive fine-tuning methods while maintaining the operational efficiency and flexibility of the Pretrain-then-Adapt paradigm. 

Our primary contributions are as follows:

\begin{figure}[!t]
    \centering
    \begin{subfigure}[b]{0.45\columnwidth}
        \centering
        \includegraphics[width=\columnwidth]{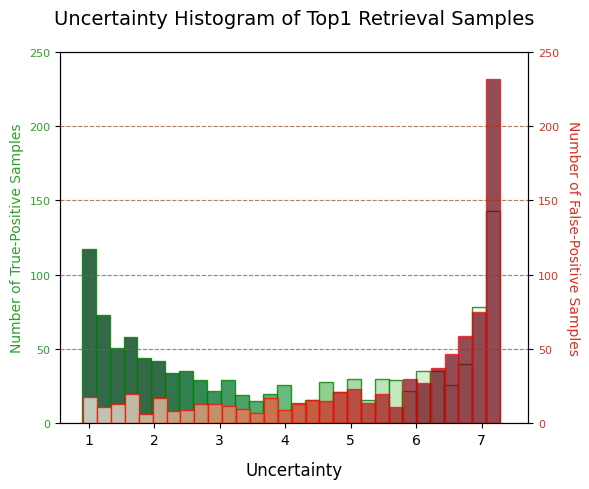}
        \caption{RSTPReid}
        \label{fig2:rstp}
    \end{subfigure}
    \hfill 
    \begin{subfigure}[b]{0.45\columnwidth}
        \centering
        \includegraphics[width=\columnwidth]{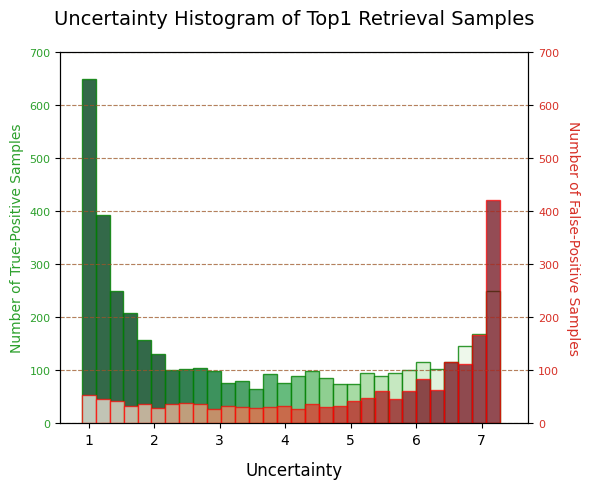}
        \caption{CUHK-PEDES}
        \label{fig2:cuhk}
    \end{subfigure}
    \vspace{-.1in}
    \caption{\textbf{Statistical Overview of our Uncertainty Indicator on (a) RSTPReid \citep{zhu2021dssl} and (b) CUHK-PEDES \citep{li2017person}}. We count the True Positives (TP) and False Positives (FP) sample number in the initial ranking list before adaptation according to the proposed uncertainty score. TP samples consistently cluster in the low-uncertainty region, while FP samples concentrate in the high-uncertainty region across both benchmarks. Therefore, it could serve as the indicator for the test-time adaptation.}
    \label{fig2}
    \vspace{-.2in}
\end{figure}

\begin{itemize}[leftmargin=2em, itemsep=0.25em]
    \item \textbf{A Practical Paradigm for Test-Time Adaptation on Text-based Person Search.} We explore a \textbf{Pretrain-then-Adapt} paradigm for text-based person search that alleviates the need for labeled target-domain data. This framework offers a practical alternative to the standard Pretrain-then-Finetune pipeline, enhancing deployability in real-world scenarios where data annotation is infeasible.

    \item \textbf{An Uncertainty-Guided Adaptation Method.} We propose an Uncertainty-Aware Test-Time Adaptation (UATTA) framework designed to address domain shift under unsupervised conditions. The method introduces a \textbf{bidirectional retrieval disagreement} mechanism to estimate prediction uncertainty. This signal is used to guide the adaptation, aiming to curb error accumulation from overconfident false positive predictions during the offline test-time optimization process.

    \item \textbf{Comprehensive Empirical Evaluation.} We conduct extensive experiments on four challenging benchmarks (CUHK-PEDES \citep{li2017person}, ICFG-PEDES \citep{ding2021seman}, RSTPReid \citep{zhu2021dssl}, and PAB \citep{yang2024beyond}). Our results show that UATTA achieves consistent performance improvements over baseline methods across different model architectures. The findings validate the efficacy of our uncertainty-guided approach and suggest it is a promising direction for label-free adaptation in this domain.
\end{itemize}

\label{1_introduction}

\section{Related Work}
\label{2_related_work}

\noindent\textbf{Text-based Person Search.}
Text-based person search aims to find the target person of interest via a text query. Different from image-based search~\citep{hou2025fire}, the text query is more intuitive for users. A typical dataset is CUHK-PEDES~\citep{li2017person}. To align person images and text, recent works usually adopt a pretrain-then-finetune paradigm, in which models first establish cross-modal alignment on synthetic person-caption data and then fine-tune on limited real-world annotations. \cite{shao2023unified} apply CLIP~\citep{radford2021learning} with a novel divide-conquer-combine strategy to automatically annotate pseudo-text descriptions for a large-scale person re-identification image dataset~\citep{fu2021unsupervised}, which reduces human labor and cost. With the help of image generative models, \citep{yang2023towards} collect a new large-scale cross-modal dataset MALS~\citep{yang2023towards}, containing real-world text descriptions and corresponding generated person images with multiple attributes, providing an alternative for real-world person privacy via automatic image generation and attribute extraction. Following this synthetic-pretrain and real-world-finetune approach, \citep{tan2024harnessing, jiang2025modeling} boost text-based person search performance by exploiting Multi-modal Large Language Models to obtain text descriptions with various language structures and styles. Existing test-time inference pipelines of this paradigm can be divided into one-stage CLIP-based~\citep{radford2021learning} and XVLM-based~\citep{zeng2021multi} frameworks. The former~\citep{cvpr23crossmodal, shao2023unified, tan2024harnessing, jiang2025modeling, chen2025class} extracts vision and language features independently via separate single-modal models and predicts alignment based on Image-Text Contrastive (ITC) similarity~\citep{radford2021learning}. The latter~\citep{li2021align, li2022blip, li2023blip, zeng2021multi, yang2023towards, qu2023learnable, yang2024beyond, su2024maca, wang2025beyond} employs an additional fine-grained cross-modal interaction module to exploit Image-Text Matching (ITM) learning and predict binary matching results to rectify top-$K$ results from the first stage. In this paper, we propose a universal Pretrain-then-Adapt paradigm that is not constrained by the scarcity of annotated labels for both one-stage and two-stage frameworks.

\noindent\textbf{Test-Time Adaptation.}
Test-time Adaptation (TTA) has emerged as a promising paradigm that dynamically aligns the model with the specific test distribution during inference, effectively mitigating domain shift without source data access.
Parameter-metric approaches~\citep{wang2021tent, yang2022test} minimize prediction entropy through lightweight parameter updates, e.g., BatchNorm~\citep{ioffe2015batch} statistics. However, these approaches suffer from confirmation bias as domain shift induces high-confidence errors, a phenomenon exacerbated in cross-modal retrieval where false positives deteriorate performance~\citep{zhao2024testtime}. Memory-based approaches~\citep{iwasawa2021test, zhang2023ada} maintain feature banks for pseudo-label refinement but introduce prohibitive computational overhead for memory indexing and require structural modifications incompatible with frozen VLM backbones. Recent works~\citep{niu2023towards, tan2025uncert, niu2025test} attempt to reduce overhead through sample selection, but these strategies focus on a small number of high-confidence samples, which induces catastrophic forgetting by overfitting and deviates from pretrained feature manifolds~\citep{lee2024entropy}. Notably, our work reformulates offline test-time adaptation through uncertainty-weighted entropy minimization on the whole test set, which suppresses overconfidence on false positives while preserving frozen VLM backbones. By leveraging global domain statistics and filtering unreliable signals via cycle consistency, our approach avoids suboptimal convergence, achieving a superior balance between accuracy and efficiency for real-world deployment.

\begin{figure*}[t!]
    \centering
    \includegraphics[width=1\textwidth]{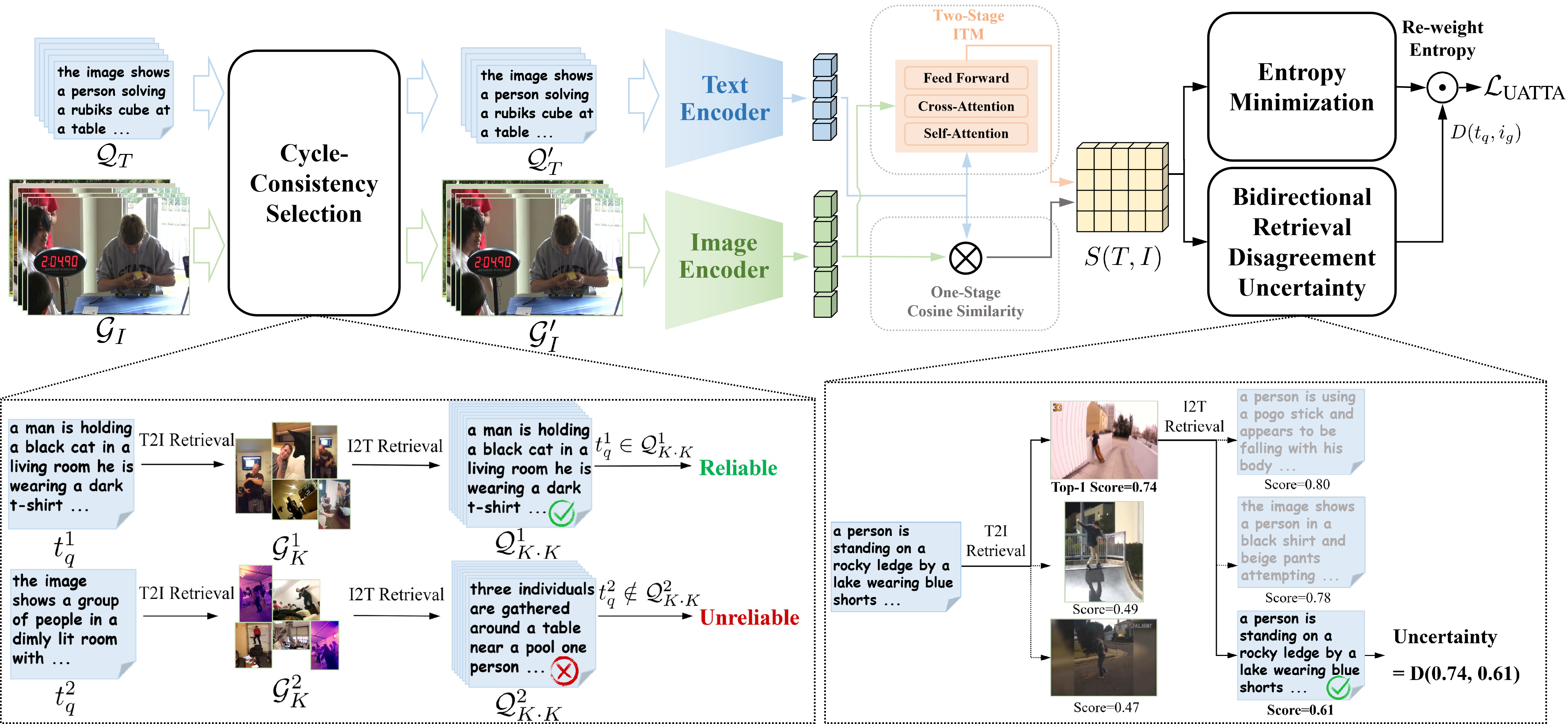}
    \vspace{-.25in}
        \caption{\textbf{Uncertainty-aware Test-Time Adaptation Framework(UATTA)}. Given the image gallery set $\mathcal{G}_I$ and text query set $\mathcal{Q}_T$ in the test set, we first select reliable samples by Cycle-Consistency Selection and obtain reliable text query set $\mathcal{Q}'_T$ and reliable image gallery set $\mathcal{G}'_I$. Then we compute the similarity matrix $S(T,I)$ for every pairs to calculate uncertainty. Finally we re-weight the entropy minimization objective with calculated uncertainty. As shown in Cycle-Consistency Selection stage, we sort samples who are mutual top-$K$ neighbors, in which the initial given text $t^1_q$ is supposed to be inversely found in $\mathcal{G}^1_{K\cdot K}$ by images in retrieval result set $\mathcal{G}^1_K$ of $t^1_q$. Conversely, text $t^2_q$ is unreliable as image set $\mathcal{G}^2_{K\cdot K}$ do not contain itself. After selection, based on the reliable images and texts, we further exploits the Bidirectional Retrieval Disagreement mechanism to estimate uncertainty with both text-to-image top-$1$ retrieval probability and inverse image-to-text retrieval probability, as detailed in Bidirectional Retrieval Disagreement Uncertainty stage. This uncertainty signal is calculated by $D(t,i)$, as detailed in Eq.~\ref{eq:newdef} and dynamically re-weights the entropy minimization objective to uncertainty-weighted gradient re-calibration loss $\mathcal{L}_{\text{UATTA}}$, as detailed in Eq.~\ref{eq:uatta_loss}.
        Our UATTA framework mitigates domain gaps with minimal adaptation cost and zero extra architecture.
        }
    \label{fig1}
\end{figure*}

\noindent\textbf{Uncertainty in Cross-Modal Retrieval.}
Uncertainty quantification has gained traction in cross-modal retrieval~\citep{chen2023rethink, wang2024semi, li2025revo}. Generally, uncertainty can be quantified as the discrepancy of representation between different modalities, which is more pronounced under domain gaps~\citep{xu2024invisible}. \citep{chen2024composed} integrate fine- and coarse-grained retrieval with different fluctuations to model uncertainty and rectify the matching objective. Furthermore,  Li~\etal~\cite{li2024adaptive} leverage subjective logic to select reliable cross-modal pairs and masked modeling to capture cross-modal relations, and also exploit multi-grained uncertainty-based alignments to mitigate domain shifts. With the help of an extra large vision-language model,  Zhao~\etal~\cite{zhao2024testtime} use CLIP to reflect the uncertainty of input pairs and boost zero-shot performance via an uncertainty-aware reward feedback mechanism. Li~\etal~\cite{li2025test} optimize the robustness of test-time adaptation via candidate selection, inter-modal gap learning, and intra-modal uniformity learning, yet are constrained to query modal shifts. Through a novel design of probabilistic distance metrics and hierarchical learning objectives, \citep{tang2025modeling} explicitly model uncertainty at multi-grained levels, enabling more nuanced and robust composed image retrieval that can handle polysemy and ambiguity in search intentions. Recent cross-modal retrieval uncertainty estimation methods, whether multi-grained or contrastive, optimize representation similarity via explicit feature-space constraints while neglecting retrieval trajectory consistency. UATTA implicitly optimizes the embedding space by leveraging the inherent consistency of correct retrieval. Specifically, our bidirectional retrieval disagreement mechanism formulates uncertainty estimation with the inherent retrieval trajectory-symmetric nature.

\section{Method}
In this section, we introduce the proposed Uncertainty-aware Test-Time Adaptation (UATTA) framework for text-based person search, as illustrated in Fig.~\ref{fig1}.
Firstly, we introduce a dynamic sample selection strategy based on the cycle-consistency to select reliable samples where the original text query can be successfully recovered. Based on the selected samples, we perform the uncertainty estimation. Finally, we integrate estimated uncertainty into test-time adaptation via entropy recalibration resulting in mitigating the adverse effects of erroneous gradients induced by overconfident false positives. It is noted that UATTA applies seamlessly in both CLIP-based one-stage and XVLM-based two-stage models.

\subsection{Similarity Matrix Generation.} 
For a given text query $t_q$, text-to-image retrieval aims to select corresponding most similar images from the image gallery set $\mathcal{G}_I=\{i_g\}_g, {g\in[1,N_I]}$. 
Within our pretrain-then-adapt paradigm, the whole text query set $\mathcal{Q}_T$ is accessible during adaptation. 
We employ the encoders of cross-modal retrieval model to map images in $\mathcal{G}_I$ and texts in $\mathcal{Q}_T$ into a shared embedding space. Subsequently, we compute the similarity scores $s(t_q,i_g)\label{subsec:SimMat}$ between pairs, forming similarity matrix $S(T,I)$. For CLIP-based one-stage retrieval models, the similarity scores is computed using cosine similarity, $s(t_q,i_g) = \cos\big(\mathcal{E}_{\text{T}}(t), \mathcal{E}_{\text{I}}(i)\big)$, where $\mathcal{E}_{\text{T}}, \mathcal{E}_{\text{I}}$ are modality-specific encoders. For XVLM-based two-stage retrieval models, the matching score $s(t_q,i_g) = \mathcal{E}_{\text{ITM}}(\mathcal{E}_{\text{T}}(t),\mathcal{E}_{\text{I}}(i))$ is obtained through an additional image-text matching (ITM) module $\mathcal{E}_{\text{ITM}}$.

\subsection{Cycle-Consistency Selection} 
\label{sec:selection}
During our pretrain-then-adapt paradigm, we first introduce the Cycle-Consistency Selection (CCS) to select reliable samples, identifying those queries that fall within the mutual top-$K$ rankings.
Given the text query $t_q$, we first retrieve the top-$K$ most similar images with it to form $\mathcal{G}_{K}$. Subsequently, for each image in $\mathcal{G}_{K}$, we perform reverse retrieval to obtain its top-$K$ text candidates. Together, these candidates form the set $\mathcal{Q}_{K\cdot K}$. We consider $t_q$ a reliable sample if and only if it is present in $\mathcal{Q}_{K\cdot K}$.
Formally, we define the reliability indicator $r(t_q) \in \{0, 1\}$ as:
\begin{equation}
    r(t_q) = 
        \begin{cases}
            1, \text{if } t_q \in \mathcal{Q}_{{K} \cdot {K}}, \\
            0, \text{if } t_q \notin \mathcal{Q}_{{K} \cdot {K}}.
        \end{cases}
\end{equation}
Then we define $\mathcal{Q}'_T = \{t'_q\}$ as the reliable text query set, where $r(t'_q)=1, t'_q \in \mathcal{Q}_{T}$, and define $\mathcal{G}'_I = \{i'_g\}$ as the reliable image gallery set, where $i'_g$ is the retrieval candidates of $t'_q$.
This sample selection retains samples with good retrieval cycle-consistency, guaranteeing them to act as reliable anchors for the subsequent adaptation, which are useful for generalization in adaptation, and discards highly inconsistent pairs, which are harmful false positives otherwise introducing detrimental noise into the optimization process. 
Interpretation of K. 

Generally, $K$ controls the trade-off between reliability and selectivity, in which a larger $K$ provides more stable cycle consistency by incorporating a broader set of candidates, while a smaller $K$ enforces stricter selection, reducing the influence of noisy matches.
Empirically, we observe that the optimal $K$ often correlates with the number of ground-truth positives per query and it can be interpreted as an approximation of the local neighborhood size in the embedding space.
We further find that performance remains stable within a reasonable range of $K$, indicating that the method is not overly sensitive to this hyperparameter.


\subsection{Bidirectional Retrieval Disagreement Uncertainty} 

\noindent \textbf{Uncertainty form a Bayesian Perspective.}
Uncertainty is typically delineated into aleatoric (data) and epistemic (model) components within a Bayesian framework~\citep{kendall2017uncertainties}. Drawing upon this taxonomy, we propose to quantify retrieval uncertainty through behavioral observation, which captures these inherent ambiguities.
We define the model's uncertainty as the variance of its parameters, $\text{Unc}(\theta) := \text{Var}(\theta)$, a standard definition from a Bayesian perspective where parameters are treated as random variables. A large $\text{Var}(\theta)$ signifies high uncertainty in the learned weights. Higher parameter variance correlates with elevated uncertainty in learned representations, enabling principled uncertainty-aware adaptation through gradient reweighting. 

However, directly computing the parameter variance is computationally intractable in deep neural networks.
To address this, we propose a tractable proxy named \textbf{Bidirectional Retrieval Disagreement}, denoted as $D(t_q,i_g)$.
We posit that the epistemic uncertainty of a retrieval model can be effectively quantified by measuring the inconsistency between its multi-modal encoders. 
Concurrently, given a pair $(t_q,i_g)$, the bidirectional retrieval disagreement is defined as the difference between the distinct text-to-image retrieval probability $p_{T2I}$ and image-to-text retrieval probability $p_{I2T}$:
\begin{equation}
    D(t_q,i_g) := \left| \left| p_{T2I}(y|t_q,i_g,\theta) - p_{I2T}(y|t_q,i_g,\theta) \right| \right|,
    \label{eq:def}
\end{equation}
$y$ is a latent matching variable, which can not be observed during adaptation, denoting whether $i_g$ and $t_q$ is truly-matched or not. Importantly, $y$ is not required in practice, and the formulation is only used for conceptual explanation. Then in this context,  $p_{T2I}(y|t_q,i_g,\theta)$, $p_{I2T}(y|t_q,i_g,\theta)$ denote the probabilities of text-to-image and image-to-text search predictions.
To operationalize this metric, we instantiate probability as temperature-scaled softmax of similarity scores, which are parameterized by the model weights $\theta$, and compute over the top-$K$ retrieved matches to focus on hard candidates, as follows:
\begin{equation}
    \begin{split}
        p_{T2I}(y|t_q,i_g) = \frac{\exp(s'(t_q,i_g) )}{\sum_{j=1}^{K} \exp(s'(t_q,i_j) )}, \\
        \quad p_{I2T}(y|t_q,i_g) = \frac{\exp(s'(i_g,t_q) )}{\sum_{j=1}^{K} \exp(s'(i_g,t_j) )},
    \end{split}
    \label{eq:prob}
\end{equation}
where $s'(t_q,i_g)$ denotes top-$K$ similarity matrix derived from $s(t_q,i_g)$, $s'(i_g, t_q)$ denotes top-$K$ similarity matrix from its transpose $s(t_q,i_g)^\top$. We adopt a standard softmax formulation with the temperature fixed to 1 (\ie no additional scaling), and thus omit the temperature term in the formulation.
While the functions $p_{T2I}$ and $p_{I2T}$ depend on different subsets of parameters (\eg, separate modal prediction modules), our analysis considers uncertainty over the entire parameter vector $\theta$. 

\noindent \textbf{Theoretical Justification.}
We now provide a theoretical sketch to justify the proportionality between the parameter variance and our proposed proxy, \textit{i.e.}, $\text{Var}(\theta) \propto D(t_q,i_g)$.
The proof is based on the principle of symmetric consistency. An idealized model with zero uncertainty ($\text{Var}(\theta) = 0$) can be represented by a single set of optimal parameters, $\theta_0 = E[\theta]$. Such a deterministic model, if well-trained, should exhibit symmetric predictions, meaning the probability of retrieving $i_g$ from $t_q$ is consistent with retrieving $t_q$ from $i_g$. Consequently, for this ideal model, the prediction disagreement is negligible:
\begin{equation}
    D(t_q,i_g)|_{\theta=\theta_0} = \left| \left| p_{T2I}(y|t_q,i_g,\theta_0) - p_{I2T}(y|t_q,i_g,\theta_0) \right| \right| \approx 0.
    \label{eq:ideal_model_assumption}
\end{equation}
In a realistic model, however, uncertainty implies that $\text{Var}(\theta) > 0$. The parameters $\theta$ are subject to perturbations around their mean $\theta_0$. These parameter perturbations disrupt the model's symmetric consistency, as they affect the distinct computational paths of $p_{T2I}$ and $p_{I2T}$ differently.

To formalize the relationship between parameter variance and prediction disagreement, we analyze the effect of these perturbations using a first-order Taylor expansion of the prediction functions around $\theta_0$ (here we omit $t$ and $i$ for simplicity):
\begin{align}
    p_{T2I}(y|\theta) &\approx p_{T2I}(y|\theta_0) + (\theta - \theta_0)^T \nabla_{\theta} p_{T2I}(y|\theta_0), \\
    p_{I2T}(y|\theta) &\approx p_{I2T}(y|\theta_0) + (\theta - \theta_0)^T \nabla_{\theta} p_{I2T}(y|\theta_0).
\end{align}
By substituting these into the definition of $D(t,i)$, we obtain:
\begin{equation}
    \begin{split}
        D(t_q,i_g) \approx &\ \big\| (p_{T2I}(y|\theta_0) - p_{I2T}(y|\theta_0)) \\
        & + (\theta - \theta_0)^T (\nabla_{\theta} p_{T2I}(y|\theta_0) - \nabla_{\theta} p_{I2T}(y|\theta_0)) \big\|.
    \end{split}
\end{equation}
Applying the symmetric consistency assumption, where $p_{T2I}(y|\theta_0) - p_{I2T}(y|\theta_0) \approx 0$, the expression simplifies to:
\begin{equation}
    D(t_q,i_g) \approx \left|\left| (\theta - \theta_0)^T (\nabla_{\theta} p_{T2I}(y|\theta_0) - \nabla_{\theta} p_{I2T}(y|\theta_0)) \right|\right|.
\end{equation}
This result demonstrates that the magnitude of the prediction disagreement $D(t_q,i_g)$ is directly dependent on the parameter deviation $(\theta - \theta_0)$. Since $\text{Var}(\theta) = E[(\theta - \theta_0)^2]$ measures the expected squared magnitude of this deviation, a larger parameter variance will lead to a larger expected prediction disagreement. This establishes the proportionality $\text{Var}(\theta) \propto D(t_q,i_g)$, validating the use of prediction disagreement as a computationally efficient and theoretically grounded proxy for model uncertainty. 


\subsection{Uncertainty-aware Test-Time Adaptation} 
From the preceding Cycle-Consistency Selection quantification procedure, we obtain a reliable subset with good retrieval cycle consistency. Leveraging this curated sample set, we perform adaptation through entropy minimization with uncertainty-weighted gradients, effectively instantiating the principle of input-dependent loss attenuation in Bayesian framework~\citep{kendall2017uncertainties}. This strategy aligns the model's feature distribution to the target domain while preserving cross-modal consistency. Consequently, we bridge the synthetic-to-real domain gap without requiring labeled target-domain supervision.
Empirically, we find that raw probability differences defined in Eq.~\ref{eq:def} are insufficient to capture bidirectional disagreement.
When both $p_{T2I}$ and $p_{I2T}$ approach zero, their absolute difference is negligible, falsely implying high consistency. Therefore, in practice, we normalize the absolute difference by the mean value to penalize low-confidence pairs while preserving consistency for high-confidence matches. Furthermore, we employ exponential amplification to accentuate the discriminative contrast between asymmetric matches (one high probability, one low) and symmetric high-confidence matches, modifying $D(t_q,i_g)$ as:
\begin{equation}
    D(t_q,i_g) := \exp\bigg(\dfrac{|p_{T2I}(y|t_q,i_g) - p_{I2T}(y|t_q,i_g)|}{\frac{p_{T2I}(y|t_q,i_g) + p_{I2T}(y|t_q,i_g)}{2}}\bigg).
    \label{eq:newdef}
\end{equation}
Normalization prevents degenerate cases where both probabilities are small, while exponential amplification enhances the contrast between asymmetric matches and symmetric high-confidence matches. 
Further ablation studies analyzing the contribution of each component are provided in Sec.~\ref{subsec:unc_formula}.
This design is consistent with common practices in uncertainty calibration, where normalization and scaling are used to improve discriminative behavior.

To date, previous TTA method\citep{wang2021tent} employs entropy minimization objective $\mathcal{L}_{\text{Tent}} = - \sum p \log(p)$ for classification adaptation, where $p$ denotes the model's prediction probability distribution, suffering from overconfident predictions on false-positive samples\citep{wang2021tent, zhao2024testtime}.
We thus far reformulate the bidirectional adaptation objective combined with Eq.~\ref{eq:newdef} through uncertainty-weighted gradient re-calibration:
\begin{equation}
    \begin{split}
        \mathcal{L}_{\text{UATTA}} =  &  \sum_{i_g \in \mathcal{G}'_I, t_q \in \mathcal{Q}'_T} \bigg( \frac{ - p_{T2I}(y|t_q,i_g) \log(p_{T2I}(y|t_q,i_g))}{D(t_q,i_g)} \\
         & + \frac{ - p_{I2T}(y|t_q,i_g) \log(p_{I2T}(y|t_q,i_g))}{D(t_q,i_g)} \bigg),
    \end{split}
    \label{eq:uatta_loss}
\end{equation}
where $\mathcal{G}'_I,\mathcal{Q}'_T$ are the reliable image gallery set and text query set obtain through Cycle-Consistency Selection as described in Sec.\ref{sec:selection}. 

\noindent \textbf{Analysis.} The Bidirectional Retrieval Disagreement $D(t_q,i_g)$ serves as an uncertainty-weighted recalibration mechanism, adaptively modulating the contribution of each text-image pair to the entropy minimization objective. 
Specifically, low-uncertainty pairs, which predominantly correspond to true positives as illustrated in Fig.~\ref{fig2}, receive amplified gradient updates that strengthen cross-modal alignment. Conversely, high-uncertainty pairs undergo gradient suppression, preventing error propagation from ambiguous or false matches. 
This dual consistency constraint, which requires cycle-consistent retrieval from both text-to-image and image-to-text directions, naturally partitions samples into confident matches and uncertain candidates without auxiliary supervision.
Remarkably, UATTA achieves effective label-free adaptation under domain shift through implicit embedding space optimization, with minimal adaptation cost and zero architectural modifications to pretrained vision-language models.

\label{3_method}

\section{Experiment}
\begin{table}[t!]
    \centering
    \caption{Quantitative comparison of our proposed \textbf{Pretrain-then-Adapt} paradigm with state-of-the-art methods on Text-based Person Anomaly Search benchmark PAB \citep{yang2024beyond}. The gpu used for post-training is NVIDIA GeForce RTX 3090 GPU. Best results are \textbf{bold}. Second best results are \underline{underlined}.}

    \label{table:pab}
    \small
    \setlength{\tabcolsep}{3pt}
    \resizebox{\columnwidth}{!}{
        \begin{tabular}{@{}l|l|c|cccc@{}}
            \toprule
            Method & Type & Post-train & R@1 & R@5 & R@10 & mAP \\
            \midrule
            \multicolumn{7}{@{}l}{\textit{Pure pretraining (no adaptation / finetuning)}} \\
            \midrule
            MRA~\citep{yang2025mini}      & Pretrain & — & 9.91  & 23.66 & 31.45 & 17.15 \\
            RaSa~\citep{bai2023rasa}      & Pretrain & — & 21.74 & 27.30 & 27.96 & 24.35 \\
            WoRA~\citep{sun2025from}      & Pretrain & — & 22.25 & 45.91 & 53.54 & 33.39 \\
            APTM~\citep{yang2023towards}  & Pretrain & — & 22.90 & 45.80 & 52.38 & 33.56 \\
            CAMeL~\citep{yu2025camel}     & Pretrain & — & 24.47 & 50.00 & 58.75 & 36.75 \\
            IRRA~\citep{cvpr23crossmodal} & Pretrain & — & 30.59 & 59.61 & 68.91 & 44.41 \\
            CLIP~\citep{radford2021learning}  & Pretrain & — & 47.57 & 81.55 & 89.03 & 62.73 \\
            X-VLM~\citep{zeng2021multi}   & Pretrain & — & 71.94 & 97.78 & 98.99 & 83.96 \\
            \midrule
            \multicolumn{7}{@{}l}{\textit{Pretrain-then-Finetune (Pre-FT)}} \\
            \midrule
            MRA~\citep{yang2025mini}      & Pre-FT & 1.06h (4 GPUs) & 70.53 & 94.69 & 97.47 & 81.59 \\
            APTM~\citep{yang2023towards}  & Pre-FT & 0.51h (4 GPUs) & 72.14 & 95.30 & 97.17 & 82.78 \\
            CAMeL~\citep{yu2025camel}     & Pre-FT & 1.01h (4 GPUs) & 74.30 & 96.79 & 98.84 & 84.20 \\
            WoRA~\citep{sun2025from}      & Pre-FT & 0.88h (4 GPUs) & 74.47 & 96.82 & 98.48 & 84.60 \\
            IRRA~\citep{cvpr23crossmodal} & Pre-FT & 19.6h (4 GPUs) & 76.39 & 97.62 & 99.14 & 86.33 \\
            CLIP~\citep{radford2021learning}  & Pre-FT & 18.4h (4 GPUs) & 77.60 & 98.84 & 99.75 & 87.35 \\
            RaSa~\citep{bai2023rasa}      & Pre-FT & 0.74h (4 GPUs) & 80.79 & 98.89 & 99.65 & 89.20 \\
            X-VLM~\citep{zeng2021multi}   & Pre-FT & 40.5h (4 GPUs) & 81.95 & 98.84 & 99.19 & 89.86 \\
            X-VLM + CMP~\citep{yang2024beyond} & Pre-FT & 48.1h (4 GPUs) & 84.93 & 99.09 & 99.75 & 91.66 \\
            \midrule
            \multicolumn{7}{@{}l}{\textit{Pretrain-then-Adapt (Pre-Adp)}} \\
            \midrule
            X-VLM + SAR~\citep{niu2023towards}  & Pre-Adp & 0.38h (1 GPU) & 73.20 & \underline{97.87} & \underline{99.09} & 84.58 \\
            X-VLM + Tent~\citep{wang2021tent}   & Pre-Adp & \underline{0.23h (1 GPU)} & 73.50 & 95.65 & 97.57 & 83.71 \\
            X-VLM + SHOT~\citep{liang2020we}    & Pre-Adp & 0.26h (1 GPU) & 73.66 & 95.80 & 97.82 & 83.97 \\
            X-VLM + READ~\citep{yang2024test}   & Pre-Adp & \underline{0.23h (1 GPU)} & 74.62 & 96.00 & 98.18 & 84.61 \\
            X-VLM + TCR~\citep{li2025test}      & Pre-Adp & 0.25h (1 GPU) & \underline{74.92} & 96.15 & 97.97 & \underline{84.72} \\
            \midrule
            \textbf{X-VLM + Ours} & Pre-Adp & \textbf{0.08h (1 GPU)} & \textbf{76.13} & \textbf{98.02} & \textbf{99.09} & \textbf{86.14} \\
            \bottomrule
        \end{tabular}
    }
    \vspace{-.15in}
\end{table}

\begin{table*}[!t]
    \centering
    \caption{Quantitative comparison of our proposed \textbf{Pretrain-then-Adapt} paradigm with state-of-the-art direct transfer models and other existing \textbf{Test-Time-Adaptation}, \textbf{Semi-supervised} and \textbf{Unsupervised} methods on real-world text-based person search benchmarks \citep{li2017person, zhu2021dssl, ding2021seman}. Best results are \textbf{bold}. Second best results are \underline{underlined}.}
    \small
    \setlength{\tabcolsep}{4pt}
    \renewcommand{\arraystretch}{1.2}
    \label{table:3bench}
    \setlength{\tabcolsep}{2.5pt}
    \begin{tabular}{@{}c *{12}{c}@{}} 
        \toprule
        \multirow{2}{*}{\textbf{Method}} & \multicolumn{4}{c}{\textbf{RSTPReid}} & \multicolumn{4}{c}{\textbf{CUHK-PEDES}} & \multicolumn{4}{c}{\textbf{ICFG-PEDES}} \\
        \cmidrule(lr){2-5} \cmidrule(lr){6-9} \cmidrule(lr){10-13}
        & R1 & R5 & R10 & mAP & R1 & R5 & R10 & mAP & R1 & R5 & R10 & mAP \\
         \midrule
         \textit{Pure pretraining (no adaptation / finetuning)} \\
        \midrule
        CLIP~\citep{radford2021learning} & 12.65 & 27.16 & -- & 11.15 & 6.67 & 17.91 & -- & 2.51 & 13.45 & 33.85 & -- & 10.31 \\
        LuPerson-T~\citep{shao2023unified} & 22.40 & -- & -- & 17.08 & 21.88 & -- & -- & 19.96 & 11.46 & -- & -- & 4.56 \\
        SYNTH-PEDES~\citep{zuo2024plip} & 42.69 & -- & -- & 31.18 & 57.58 & -- & -- & 52.45 & 57.08 & -- & -- & 32.06 \\
        LuPerson-MLLM~\citep{tan2024harnessing} & 51.65 & 74.20 & 82.85 & 38.31 & 38.29 & 56.60 & 64.56 & 20.43 & 57.61 & 75.99 & 82.76 & 51.45 \\
        LuPerson-HAM~\citep{jiang2025modeling} & 59.50 & 80.05 & 87.05 & 44.11 & 70.59 & 86.89 & 91.78 & 63.39 & 60.64 & \textbf{77.50} & \textbf{83.26} & \underline{35.54} \\
        \midrule
        \textit{Unsupervised Domain Adaptation} \\
        \midrule
        GAAP~\citep{li2024cross} & 44.45 & 65.15 & 75.30 & 31.21 & 47.64 & 67.79 & 76.08 & 41.28 & 27.12 & 44.91 & 53.56 & 11.43 \\
        GTR~\citep{bai2023text} & 46.65 & 70.70 & 80.65 & 34.95 & 48.49 & 68.88 & 76.51 & 43.67 & 29.64 & 47.23 & 55.54 & 14.20 \\
        PSPD~\citep{chen2025unsuper} & 48.50 & 69.95 & 78.50 & 34.83 & 53.47 & 72.81 & 76.57 & 46.41 & 38.49 & 53.40 & 60.35 & 16.49 \\
        MUMA~\citep{li2025exploring} & 54.35 & 76.05 & 83.65 & 40.50 & 59.52 & 77.79 & 84.65 & 52.75 & 38.11 & 56.01 & 63.96 & 19.02 \\
        \midrule
        \textit{Semi-supervised Domain Adaptation} \\
        \midrule
        CMMT~\citep{zhao2021weakly} & -- & -- & -- & -- & 57.10 & 78.14 & 85.23 & -- & -- & -- & -- & -- \\
        Generation-then-Retrieval~\citep{gao2025semi} & 56.45 & -- & -- & 44.45 & 63.87 & -- & -- & 57.18 & 46.46 & -- & -- & 26.90 \\
        TextReID~\citep{han2021textreid} & -- & -- & -- & -- & 64.40 & 81.27 & 87.96 & 61.19 & -- & -- & -- & -- \\
        ECCA~\citep{gong2024enhanc} & -- & -- & -- & -- & 68.13 & \textbf{87.26} & 91.88 & - & -- & -- & -- & -- \\
        \midrule
         \textit{Pretrain-then-Adapt (Pre-Adp)} \\
        \midrule
        LuPerson-HAM + CoOp~\citep{zhou2022learning} & 58.60 & 79.65 & 87.50 & 43.65 & 70.09 & 86.48 & 91.32 & 63.10 & 60.28 & 76.24 & 82.31 & 35.16 \\
        LuPerson-HAM + SAR~\citep{niu2023towards} & 59.55 & 80.05 & 87.00 & 44.12 & 70.63 & 86.87 & 91.79 & 63.40 & \underline{60.64} & \textbf{77.50} & \underline{83.25} & \underline{35.54} \\
        LuPerson-HAM + Tent~\citep{wang2021tent} & 59.65 & 79.75 & 87.30 & 44.24 & 70.30 & 87.02 & 91.74 & 63.26 & 59.59 & 76.89 & 82.85 & 34.86 \\
        LuPerson-HAM + READ~\citep{yang2024test} & 59.80 & 79.90 & 87.30 & 44.37 & 70.06 & 86.98 & 91.82 & 63.12 & 60.31 & 77.09 & 82.96 & 35.27 \\
        LuPerson-HAM + SHOT~\citep{liang2020we} & 60.10 & 79.85 & 87.10 & 44.46 & 70.43 & 86.90 & \underline{91.99} & 63.30 & 60.31 & 76.95 & 82.86 & 35.10 \\
        LuPerson-HAM + TCR~\citep{li2025test} & \underline{61.00} & \underline{80.85} & \underline{88.35} & \underline{45.94} & \underline{70.66} & \underline{87.21} & \textbf{92.13} & \textbf{63.60} & 59.32 & 75.63 & 81.63 & 35.13 \\
        LuPerson-HAM + \textbf{Ours} & \textbf{61.85} & \textbf{81.40} & \textbf{88.40} & \textbf{46.37} & \textbf{70.92} & 86.89 & 91.86 & \underline{63.50} & \textbf{62.15} & \underline{77.31} & 82.95 & \textbf{36.11} \\
        \bottomrule
    \end{tabular}
\end{table*}

\subsection{Experiment Setting} 
We conduct experiments on two distinct frameworks for text-based person search: a one-stage retrieval framework and a two-stage retrieve-and-match framework. These choices allow us to evaluate our approach on tasks with varying complexity, from standard retrieval to fine-grained matching. \textbf{(1) CLIP-based One-Stage Framework.} For the standard person retrieval task, we adopt the state-of-the-art LuPerson-HAM model as our baseline. Our experiments are conducted on three real-world benchmarks: RSTPReid, CUHK-PEDES, and ICFG-PEDES. A key challenge is that LuPerson-HAM is pre-trained on synthetic annotations, which creates a significant domain gap compared to the human-annotated captions in the test sets. Our test-time adaptation method is designed to bridge this gap. \textbf{(2) XVLM-based Two-Stage Framework.} For the more complex person anomaly search task, which requires both coarse-grained retrieval and fine-grained matching, we follow the state-of-the-art CMP model. This model, based on the XVLM architecture, is evaluated on the PAB benchmark. Similar to the one-stage setup, PAB's training data is synthetically generated, while its test data consists of real-world images with human-corrected captions, presenting a clear domain gap that motivates our approach.

\noindent \textbf{Implementation Details.} During test-time adaptation, we optimize only the affine parameters ($\gamma$ and $\beta$) of the Layer Normalization layers within the final six layers of the text encoder. This specific choice is made to maintain consistency with the CMP baseline, where these last six layers correspond to the cross-modal attention blocks essential for image-text matching. We adopt the AdamW optimizer for all experiments. For the LuPerson-HAM baseline, the learning rate is set to $1e-3$, with a query texts number of 32 and a positive-to-negative image sample ratio of 1:3. 
For the XVLM baseline, the learning rate is $1e-4$, the number of query texts is 16, and the sample ratio is 1:7. The batch size is maintained at a constant 128, configured jointly by the number of query texts and the specified positive-to-negative ratio. The number of adaptation rounds is adjusted based on the test set size, \ie, 50 for PAB and RSTP-Reid, and 10 for ICFG-PEDES and CUHK-PEDES.

\subsection{Comparison with State-of-the-arts}

\noindent\textbf{Comparison with Pretrain Models.}
We compare our method with state-of-the-art methods on multiple benchmarks. As shown in Table~\ref{table:pab}, our method significantly improves +4.19\% R@1 compared to pretrained XVLM, which proves the capacity of our Pretrain-then-Adapt paradigm on mitigating domain gaps between unrelated pretrained data and specific person anomaly search data. 
Notably, our pretrain-then-adapt paradigm achieves significant efficiency gains with merely 0.08 hours of adaptation time. The process of adaptation operates directly on unlabeled test data of target domain, while others need finetuning on labeled train data of target domain, consuming additional post-train burden. Although some models, benefiting from lightweight fine-tuning modules, reduce post-train time from dozens of hours to approximately one hour, they still require 4 NVIDIA GeForce RTX 3090 GPU whereas only single 3090 GPU for ours. The efficiency gains become particularly significant when considering practical deployment constraints in privacy-sensitive and resource-constrained environments.
We observe a similar improvement on three text-based person search benchmarks in Table~\ref{table:3bench}. 
The results show that the R@1 score increases 2.35\%, 0.33\% and 1.51\% on RSTPReid, CUHK-PEDES and ICFG-PEDES respectively, and the mAP score is improved by 2.26, 0.29 and 0.57. These boosts underscore the efficacy of our proposed bidirectional retrieval disagreement uncertainty and sample selection in mitigating the impact from false positives, which generally refines model to be overconfident in traditional entropy minimization test-time adaptation methods.

\begin{table}[!t]
    \centering
    \caption{Comparison of \textbf{Uncertainty Formulations} on RSTPReid \citep{zhu2021dssl} benchmark. 
    $p_{\text{T2I}}$ is the text-to-image retrieval probability, $p_{\text{I2T}}$ is the inverse retrieval probability, which uses the gallery image from $p_{\text{T2I}}$ to retrieve the corresponding query text. $N_{\text{T2I}}$ is the size of image gallery per identity. $N_{\text{I2T}}$ is the size of text query per identity. $\epsilon$ is a small constant to prevent divided by zero. $\mathbf{s}^{\text{top-$K$}}_{\text{T2I}}$ denotes the top-$K$ similarity matrix of text-to-image retrieval. Equally, $\mathbf{s}^{\text{top-$K$}}_{\text{I2T}}$ denotes the top-$K$ similarity matrix of inverse directional image-to-text retrieval. The similarity matrix is transformed by a softmax function to obtain retrieval probabilities for the top-$K$ results. Best results are \textbf{bolded}.}
    \label{table:ablation_unc}
    \footnotesize
    \setlength{\tabcolsep}{2pt} 
    \renewcommand{\arraystretch}{1.2}
    \begin{tabular}{>{\raggedright\arraybackslash}p{3.3cm} >{\raggedright\arraybackslash}m{2.3cm} *{4}{S[table-format=2.2]}}
        \toprule
        \multirow{2}{*}{\textbf{Uncertainty Formulation}} & 
        \textbf{Bidirectional} & 
        \multicolumn{4}{c}{\textbf{RSTPReid}} \\
        \cmidrule(lr){3-6}
        & \textbf{Retrieval Probability} & \textbf{R1} & \textbf{R5} & \textbf{R10} & \textbf{mAP} \\
        \midrule
       
        $\exp\left(1 - \dfrac{p_{\text{T2I}} + p_{\text{I2T}}}{2}\right)$ & 
        $p_{\text{T2I}} = \text{softmax}(\mathbf{s}^{\text{top-$K$}}_{\text{T2I}})$
        $p_{\text{I2T}} = \text{softmax}(\mathbf{s}^{\text{top-$K$}}_{\text{I2T}})$
        & 61.40 & 80.50 & 87.75 & 46.16 \\
        \addlinespace
        \hline
     
        $\left|\log(p_{\text{T2I}} + \varepsilon) - \log(p_{\text{I2T}} + \varepsilon)\right|$ & 
        same as above 
        & 61.75 & 81.30 & 88.40 & 45.88 \\
        \addlinespace
        \hline

        $\exp\left(\dfrac{|p_{\text{T2I}} - p_{\text{I2T}}|}{\frac{p_{\text{T2I}} + p_{\text{I2T}}}{2}}\right)$ & 
         same as above 
        & \textbf{61.85} & 81.40 & 88.40 & 46.37 \\
        \addlinespace
\hline

        $\exp\left(\dfrac{|p_{\text{T2I}} \cdot N_{\text{T2I}} - p_{\text{I2T}} \cdot N_{\text{I2T}} |}{\frac{ p_{\text{T2I}} \cdot N_{\text{T2I}} + p_{\text{I2T}} \cdot N_{\text{I2T}} } {2} }\right)$ & 
         same as above 
        & 61.75 & 81.35 & 88.60 & \textbf{46.58} \\
        \addlinespace
\hline

        $\exp\left(\dfrac{|p_{\text{T2I}} - p_{\text{I2T}} \cdot \frac{N_{\text{I2T}}}{N_{\text{T2I}}}|}{\frac{p_{\text{T2I}} + p_{\text{I2T}} \cdot \frac{N_{\text{I2T}}}{N_{\text{T2I}}}}{2}}\right)$ &
         same as above
        & 61.70 & 81.45 & 88.60 & 46.57 \\
        \addlinespace
 \hline
        
        $\exp\left(\dfrac{|p_{\text{T2I}} \cdot N_{\text{T2I}} - p_{\text{I2T}} \cdot N_{\text{I2T}}|}{\frac{p_{\text{T2I}} \cdot N_{\text{T2I}} + p_{\text{I2T}} \cdot N_{\text{I2T}}}{2}}\right)$ & 
        $p_{\text{T2I}} = \text{softmax}(\mathbf{s}^{\text{top-$K$}}_{\text{T2I}} \cdot N_{\text{T2I}})$
        $p_{\text{I2T}} = \text{softmax}(\mathbf{s}^{\text{top-$K$}}_{\text{I2T}} \cdot N_{\text{I2T}})$ 
        & 61.75 & \textbf{81.90} & \textbf{88.90} & 46.47 \\
        \addlinespace
    \hline     

        $\exp\left(\dfrac{|p_{\text{T2I}} / N_{\text{T2I}} - p_{\text{I2T}} / N_{\text{I2T}}|}{\frac{p_{\text{T2I}} / N_{\text{T2I}} + p_{\text{I2T}} / N_{\text{I2T}}}{2}}\right)$ & 
         same as above 
        & 61.30 & 81.40 & 88.55 & 46.04 \\
        \bottomrule
    \end{tabular}
\end{table}

\noindent\textbf{Comparison with other TTA methods.} We modify others test-time adaptation methods, \ie, Tent\citep{wang2021tent}, SHOT\citep{liang2020we}, SAR\citep{niu2023towards}, READ\citep{yang2024test}, TCR\citep{li2025test} from fully Test-Time Adaptation paradigm\citep{wang2021tent} to our Pretrain-then-Adapt paradigm on RSTPReid, CUHK-PEDES and ICFG-PEDES.
As shown in Table~\ref{table:pab}, Our method demonstrates superior performance and efficiency, achieving gains of 1.21\% in R@1 and 1.42\% in mAP over all compared baselines, with 0.15 fewer hours of adaptation time on 1 gpu.
As shown in Table~\ref{table:3bench}, it is evident on ICFG-PEDES that all test-time adaptation methods fail and our method outperforms baseline by 1.51\% R@1 and 0.57\% mAP, but our method also has performance degradation at R@5 and R@10, because the bidirectional retrieval disagreement mechanism is designed to rectify the harmfulness from top-1 false positives and neglects possible potential true positives in top-2 to top-10 range. This is a future direction for us to explore smooth utilization of these potential true positives in edge zone. Similar situation occurs on CUHK-PEDES, as our method achieves best R@1 of 70.92\% but is inferior to TCR on R@5, R@10 and mAP. On RSTPReid, our method surpasses other existing methods.

\noindent\textbf{Comparison with other Semi-supervised and Unsupervised Methods.}
Generally, unsupervised~\citep{li2025exploring,chen2025unsuper,bai2023text,li2024cross} and semi-supervised~\citep{gao2025semi,han2021textreid,zhao2021weakly,gong2024enhanc} paradigms for text-based person search leverage advanced VLMs to synthesize pseudo-annotations, serving as proxies for supervised image-text pairs. However, this reliance on synthetic data inevitably introduces intrinsic domain shifts. In contrast, our approach performs direct adaptation on the test data. Despite the absence of ground-truth pairings, the textual descriptions remain aligned with the target domain. Consequently, our method focuses on mitigating the distribution shifts of the pretrained model, avoiding the noisy discriminative supervision characteristic of prior approaches. As evidenced in Table~\ref{table:3bench}, existing unsupervised and semi-supervised methods struggle to fully leverage the pretrained model's capacity, often compromising representation quality due to label noise, thereby limiting their practical generalization potential in complex real-world deployment scenarios.

\subsection{Ablation Studies and Further Discussion}

\noindent\textbf{Effect of Uncertainty Formulation.}
\label{subsec:unc_formula}
We present an ablation study on the formulation of uncertainty in Table~\ref{table:ablation_unc}. The core thought is to assign a lower uncertainty on both top retrieval directions and a higher uncertainty while only uni-directional retrieval works. Formulation 1 in Table~\ref{table:ablation_unc} only considers the higher similarity of true positives but ignore the difference between TP and FP. At a opposite perspective, formulation 2 focuses on the difference neglecting the absolute numerical magnitude. Combining with two views, formulation 3 achieves best score on RSTPReid, while the others are scaled version based on formulation 3 to balance the number of positive samples in the two retrieval directions. The extreme amplifications and balances destroy the suitable consistent distribution of TP and FP, then consequently weaken performance at R@1 score, which is the primary standard we use to choose formulation.

\begin{figure}[t!]
    \centering
    \includegraphics[width=1\columnwidth]{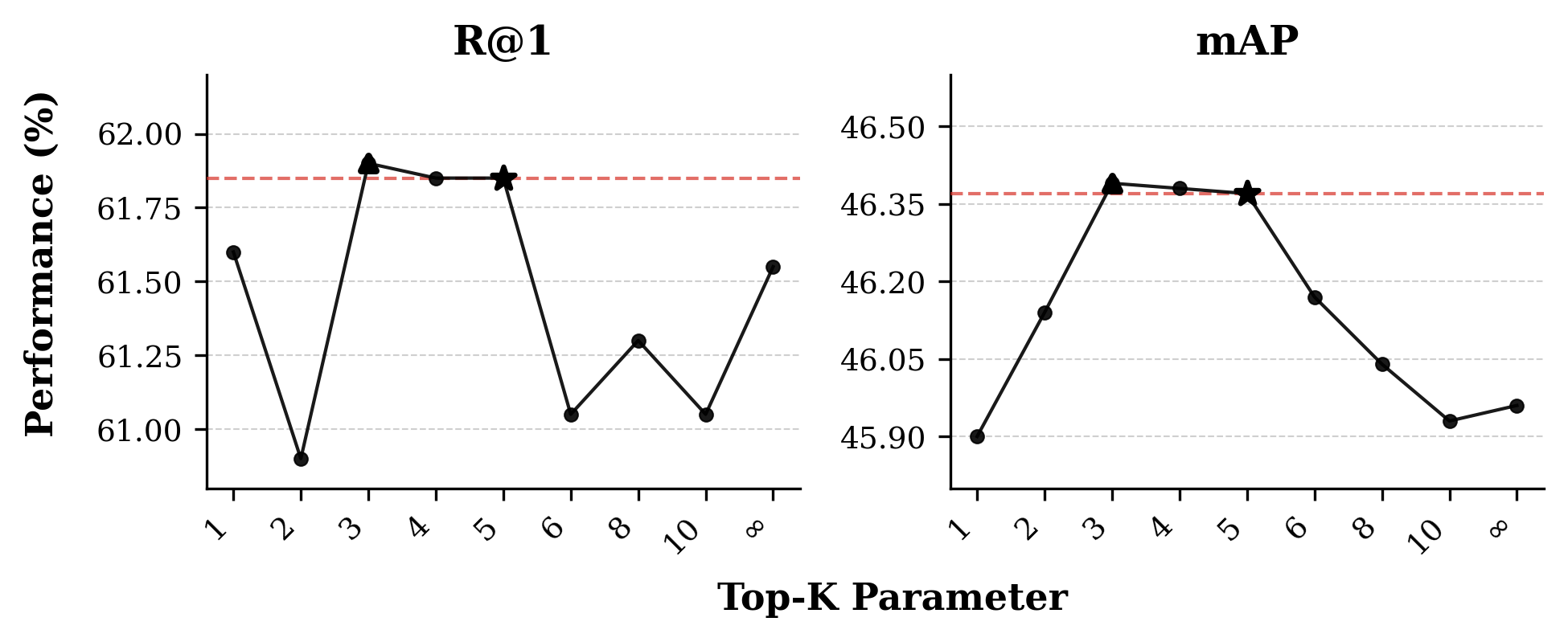}
    \vspace{-.3in}
    \caption{
        Ablation study of bidirectional Top-$K$ retrieval consistent sample selection on RSTPReid.
        $K$ denotes the mutual top range in bidirectional retrieval.
        Best performance is achieved at $K=3$.
        Since each identity in RSTPReid contains 5 ground-truth images, we adopt $K=5$ as the default setting to represent the borderline of true and false positives.
        }
    \label{fig:ablation_more_topK}
    \vspace{-.15in}
\end{figure}

\noindent\textbf{Effect of $K$ Mutual Neighbours.} 
We conduct an ablation study on the hyper-parameter $K$ in the Cycle-Consistency Selection, as shown in Fig.~\ref{fig:ablation_more_topK}. 
Based on empirical results, we adopt $K=5$ as the default setting without dataset-specific tuning, which achieves a favorable balance between performance and stability.

Specifically, smaller values such as $K=1$ restrict the adaptation process to only highly confident image-text pairs, limiting the diversity of selected samples and reducing generalization ability. 
In contrast, larger values (e.g., $K=\infty$) include all candidate pairs without selection, introducing a substantial number of false positives with high uncertainty, which negatively impacts adaptation performance. 
Intermediate values of $K$ allow the model to incorporate both low-uncertainty pairs and moderately uncertain pairs, enabling uncertainty to play an effective role in modulating the adaptation process.

From a methodological perspective, $K$ controls the locality of cycle-consistency: smaller $K$ enforces stricter agreement, while larger $K$ allows greater tolerance to retrieval noise. 
From a geometric viewpoint, $K$ can be interpreted as approximating the size of the local neighborhood in the embedding space. 
Empirically, we observe that the optimal $K$ correlates with the number of semantically similar instances per query, and values within the range $K \in [3, 8]$ consistently yield stable performance across datasets. 
Compared to the baseline performance of 58.50\% in R@1 and 44.11\% in mAP, our method consistently improves performance across all choices of $K$. 
Specifically, R@1 varies only within a narrow range of 60.90\% to 61.90\% (a fluctuation of 1.0\% absolute point), and mAP varies from 45.90\% to 46.40\% (a fluctuation of 0.5\% absolute point). 
Notably, this variation is significantly smaller than the overall performance gain over the baseline (+2.40\% to +3.40\% in R@1 and +1.79\% to +2.29\% in mAP), indicating that the improvement is robust and not sensitive to the specific choice of $K$. 
Importantly, our method is not sensitive to the exact choice of $K$ within this range, as demonstrated in Fig.~\ref{fig:ablation_more_topK}. 
This further supports that the choice of $K$ does not require careful tuning in practice.
Although the ablation is conducted on RSTPReid, we apply the same default setting across all datasets and observe consistent performance improvements, suggesting that the choice of $K$ generalizes well in practice.
This behavior can be attributed to the fact that the local neighborhood structure in the embedding space is relatively stable across datasets. The effectiveness of moderate $K$ values is also consistent with our uncertainty formulation, which benefits from a balance between confident and moderately uncertain pairs.

\begin{table}[!t]
    \centering
    \vspace{-.1in}
    \caption{
    \textbf{Ablation of the ratio between positive and negatives on RSTPReid benchmarks..} We apply different ratio of positive and negatives to compute entropy. The ratio of 1 : 3 improves the stability in test-time adaptation. Our default setting is in \colorbox{gray!20}{gray}. }
    \label{table:ablation_B}
    \vspace{-.05in}
    \small
    \setlength{\tabcolsep}{6pt}
    \renewcommand{\arraystretch}{1.2}
    \begin{tabular}{lrrrr}
        \toprule
        Ratio & R@1 & R@5 & R@10 & mAP \\
        \midrule
        1 : 1 & 60.70 & 81.20 & 88.50 & 46.01 \\
        \rowcolor{gray!20}
        1 : 3 & \textbf{61.85} & \textbf{81.40} & 88.40 & \textbf{46.37} \\
        1 : 5 & 60.95 & 80.85 & 88.05 & 46.10 \\
        1 : 7 & 61.55 & 81.35 & \textbf{88.75} & 46.32 \\
        1 : 15 & 60.80 & 81.00 & 88.50 & 46.24 \\
        \bottomrule
    \end{tabular}
    \vspace{-.1in}
\end{table}

\noindent\textbf{Effect of Negative Samples.} In Table~\ref{table:ablation_B}, we compare several experiments of the ratio between positive and negatives for one query. The optimal performance is presented with configuration of 1 : 3 on RSTPReid. This suggests that a suitable choice of ratio enhances the adaptation process with \text{softmax} entropy based on a formulation akin to multiple classification.

    


\begin{table}[!t]
    \centering
    \caption{Comparison of other prevailing lightweight tuning methods on RSTPReid\citep{zhu2021dssl} benchmark.}
    \label{table:ablation_peft}
    \footnotesize 
    \setlength{\tabcolsep}{2.5pt} 
    \renewcommand{\arraystretch}{1.2}
    \begin{tabular}{l|l|cccc} 
        \toprule
        Method & Tuning Layers & R@1 & R@5 & R@10 & mAP \\
        \midrule
        Baseline & -- & 60.64 & 77.50 & 83.26 & 35.54 \\
        CoOp*\citep{zhou2022learning} & Prompt Emb. & 58.60 & 79.65 & 87.50 & 43.65 \\ 
        Prefix-Tuning*\citep{li2021prefix} & Prefix Emb. & 26.25 & 51.45 & 63.75 & 23.14 \\
        LoRA*\citep{hu2022lora} & LoRA Matrix & 49.80 & 73.80 & 82.70 & 38.61 \\
        \textbf{Ours} & Norm. Layer & \textbf{61.85} & \textbf{81.40} & \textbf{88.40} & \textbf{46.37} \\ 
        \bottomrule
    \end{tabular}
    \vspace{-.1in}
\end{table}
\noindent\textbf{Comparison with Lightweight Tuning Methods.} We compare our method with lightweight tuning methods in Table~\ref{table:ablation_peft}. Baseline is LuPerson-HAM~\citep{jiang2025modeling} and * means that we try different hyperparameters, \ie, learning rate, number of virtual tokens, rank of LoRA\citep{hu2022lora} etc., for lightweight tuning methods and selected the best result. CoOp\citep{zhou2022learning}, which is a prompt learning method and belong to few-shot learning, fails with adaptation objective of entropy minimization. This failure suggests that learnable prompt tokens require labeled data to be grounded in a semantically meaningful embedding space mimicking natural language. In the absence of supervision, the adaptation process merely adjusts the cross-modal feature distribution while disregarding the intrinsic semantic representation. Additionally, Parameter Efficient Fine-Tuning (PEFT)~\citep{han2024parameter} provides a practical solution by efficiently adjusting the large models over the various downstream tasks. We also evaluated two representative PEFT methods, \ie, Prefix-Tuning~\citep{li2021prefix} and LoRA~\citep{hu2022lora}, for test-time adaptation on the RSTPReid benchmark, however, these approaches proved ineffective in our experiments. Although the trainable parameters in PEFT are lightweight, Entropy Minimization fails to provide sufficient supervision for learning discriminative representations.

\subsection{Qualitative Results}
\begin{figure}[!t]
    
    \centering
    \includegraphics[width=1.0\columnwidth]{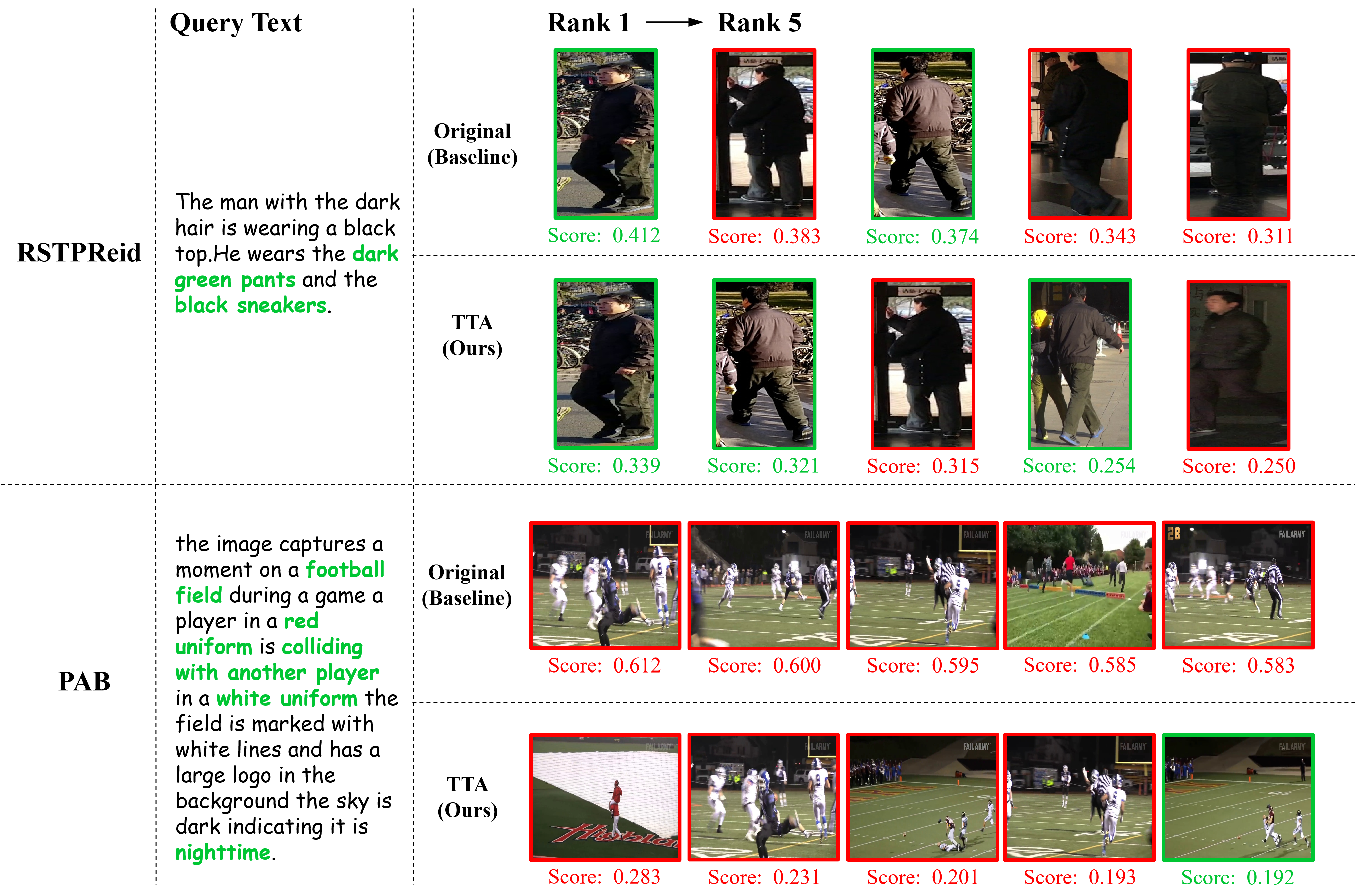}
    
    \vspace{-.15in}
    \caption{
        \textbf{Top-$5$ Text-based Person Search Results on RSTPReid and PAB.} The figure presents the Top-$5$ retrieval results for representative text queries on the RSTPReid and PAB, where the similarity score of each retrieved image is reported below the corresponding result. Correctly matched person images are highlighted with \textcolor{green}{green} bounding boxes, while false matches are indicated in \textcolor{red}{red}. 
        On RSTPReid, our method consistently promotes more ground-truth matches to higher ranks, demonstrating improved ranking quality under the text-to-image retrieval setting. In contrast, results on PAB illustrate that our approach effectively mitigates overconfident false positives by re-calibrating retrieval scores, thereby recovering correct matches that are suppressed by the baseline. These observations highlight the robustness of the proposed UATTA across different dataset characteristics.
    }
    \vspace{-.1in}
    \label{fig:vis_qual}
\end{figure}

    
    
    

\noindent\textbf{Qualitative Analysis of Person Search Performance.} To qualitatively validate the effectiveness of our Uncertainty-Aware Test-Time Adaptation (UATTA), we present a visual comparison of retrieval results between the Baseline and UATTA on the RSTPReid and PAB benchmarks in Fig.~\ref{fig:vis_qual}.
The visualization effectively showcases two key strengths of UATTA: Firstly, in challenging cases on RSTPReid where the Baseline fails due to overly high confidence in false positives, UATTA successfully rectifies the score distribution by mitigating this over-confidence, leading to the correct identification of the ground-truth image. Secondly, for scenarios requiring fine-grained semantic distinction on PAB, UATTA leverages the bidirectional retrieval disagreement proxy to effectively disambiguate subtle differences between the text and image modalities. This mechanism allows UATTA to promote more ground-truth matches to higher ranks. Overall, the qualitative results confirm that UATTA achieves robust and accurate confidence distribution by re-calibrating retrieval scores, validating its superiority in handling both retrieval ambiguity and fine-grained visual differences across different dataset characteristics.

\begin{figure}[!t]
    
    \centering
    \includegraphics[width=1.0\columnwidth]{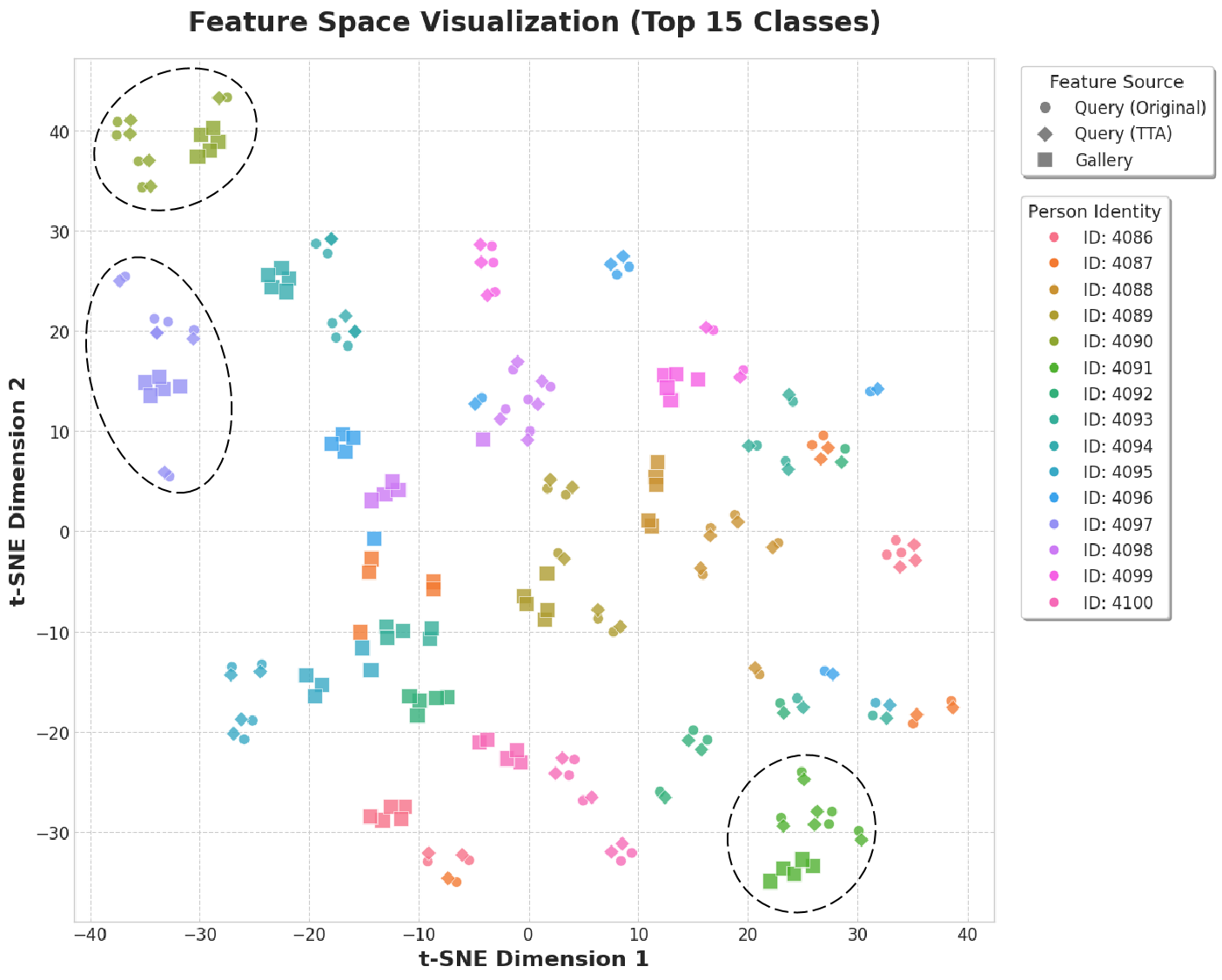}
    
    \vspace{-.15in}
    \caption{
        \textbf{T-SNE Visualization of Feature Space Shifts on RSTPReid.} 3 distinct point types represent: original query features (circles) before TTA, query features after TTA (diamonds), and gallery features (squares). Different colors distinguish individual person identities.
    }
    \vspace{-.15in}
    \label{fig:vis_tsne}
\end{figure}

\noindent\textbf{Visualization of Feature Space Shifts.} In Fig.~\ref{fig:vis_tsne}, T-SNE visualization provides an intuitive illustration of the impact of Test-Time Adaptation (TTA) on Feature Space. The visualization is focused on a representative subset of the Top-$15$ most frequent person identities to ensure clarity and showcase the adaptation effects vividly. We notice that the initial spread of original Query features (circles) demonstrates the significant domain gap and feature ambiguity present before adaptation, justifying the necessity of TTA.
After TTA, regions circled by dotted ellipses indicate that query features, post-TTA (diamonds), are effectively adapted to align more closely with their respective gallery feature (squares) clusters. 
This convergence demonstrates the efficacy of TTA in reducing feature disparity and enhancing matching performance. While the majority of person identities show strong alignment, we observe that some identities still exhibit residual ambiguity after TTA, suggesting potential avenues for future improvement in feature consolidation.


\subsection{Computational Cost Analysis}

\noindent\textbf{Complexity.}
The dominant cost comes from computing the similarity matrix $S(T, I)$, which is $\mathbf{O}(|T||I|)$, required by all retrieval baselines.
Bidirectional retrieval introduces only a constant-factor overhead (2× similarity lookup), without extra feature encoding. Since all image and text embeddings are precomputed, reverse retrieval does not require additional forward passes and feature extraction. Therefore, the overhead is negligible compared to feature encoding.

\noindent\textbf{Memory.}
Since our Pretrain-then-Adapt paradigm performs in an offline test-time adaptation manner which belongs to transductive learning setting. The similarity matrix and Cycle-Consistency Selection are precomputed inevitably on the entire dataset once time before the adaptation, which needs additional memory overhead according to the scale of different dataset. However, in the practical adaptation process, we only employs a limited number of positive and negative matches as shown in Table\ref{table:ablation_B} in a batched manner, avoiding full materialization in memory.

\label{4_experiment}

\section{Discussion and Conclusion}

In this work, we introduce a practical and label-free Pretrain-then-Adapt paradigm for text-based person search. 
We propose Uncertainty-Aware Test-Time Adaptation (UATTA), which leverages unlabeled test data to recalibrate predictions under domain shift. 
Its core component, Bidirectional Retrieval Disagreement (BRD), estimates uncertainty via discrepancies between text-to-image and image-to-text retrieval probabilities, effectively suppressing overconfident false positives while preserving reliable alignments. 
Extensive experiments on four benchmarks and both CLIP-based and XVLM-based architectures demonstrate consistent performance gains without requiring target-domain annotations or architectural changes.

\noindent\textbf{Robustness under domain shift and noisy samples.}
Under ambiguous text or low-quality images, both retrieval directions may become uniformly uncertain, reducing alignment reliability. 
In such cases, Cycle-Consistency Selection (CCS) may discard hard but correct samples, reflecting a trade-off between noise reduction and sample coverage, and partially explaining the drop in R@5 and R@10. 
Nevertheless, uncertainty-aware entropy re-calibration mitigates this issue by suppressing unreliable updates, improving robustness under moderate domain shifts.

\noindent\textbf{Self-consistency vs. correctness.}
Bidirectional Retrieval Disagreement (BRD) measures model self-consistency rather than correctness and relies on a near-deterministic pretrained model assumption. 
Our analysis, based on a first-order Taylor approximation, provides an intuitive rather than rigorous guarantee and is validated empirically. 
The method may fail in consistent-but-wrong scenarios caused by spurious correlations or calibration shifts, which require advances in domain-robust representation learning.

Overall, these limitations define the boundary of applicability but do not affect our main conclusion: uncertainty-aware test-time adaptation is an effective and efficient solution for label-free deployment under realistic domain shifts.

\label{5_conclusion_discussion}

\section{Acknowledgement} 
We acknowledge supports from Guangdong Basic and Applied Basic Research Foundation 2025A1515012281, the Jiangsu Provincial Science and Technology Program (Grant No. SBZ20250900116), the University of Macau MYRG-GRG2024-00077-FST-UMDF, and the Macao Science and Technology Development Fund Grant FDCT\slash0043\slash2025\slash RIA1.


\bibliographystyle{ACM-Reference-Format}
\balance
\bibliography{sigir2026/iclr2026_conference}


\end{document}